%% file: grbrev97.tex
\documentstyle[mprocl,psfig]{article}

\input{psfig}
\def\Journal#1#2#3#4{{#1} {\bf #2}, #3 (#4)}

\def\APJ{\em Ap. J.}
\def\APJL{\em Ap. J. Lett.}
\def\APJS{\em Ap. J. Supp.}
\def\MNRAS{\em M. N. R. A. S.}

\def\PRL{\em Phys. Rev. Lett.}
\def\etal{{\it et al.}}

\begin{document}

\title{THE GAMMA-RAY BURST MYSTERY}

\author{DAVID L. BAND}

\address{CASS, UC San Diego, La Jolla, CA 92093 USA}

\maketitle
\abstracts{
Gamma-ray bursts are transient events from beyond the solar system.
Besides the allure of their mysterious origin, bursts are physically
fascinating because they undoubtedly require exotic physics.
Optical transients coincident with burst positions show that some, and
probably all, bursts originate at cosmological distances, and not from
a large Galactic halo. Observations of these events' spectral and
temporal behavior will guide and constrain the study of the physical
processes producing this extragalactic phenomenon.} 
  
\section{Introduction}
The mystery of gamma-ray bursts and its possible solution are textbook
examples of the scientific method.  These flashes of gamma-rays
originating outside the solar system were attributed after their
discovery~\cite{kle73} to an impulsive release of energy on nearby
neutron stars.  This hypothesis had testable consequences, which the
Burst and Transient Source Experiment (BATSE) on the {\it Compton
Gamma Ray Observatory} was built to verify.\cite{fis89}  But BATSE
found the predictions were wrong, falsifying the hypothesis. Two new
hypotheses were formulated:  bursts originate in a large halo
surrounding our galaxy,\cite{lam95} or at cosmological
distances.\cite{pac95} These hypotheses also had testable
consequences, and in May, 1997, a clear signature of the cosmological
origin of at least one burst was discovered.  Other predictions of the
simplest cosmological model appear to be invalid, indicating that the
phenomenon is more complicated, and therefore more interesting, than
previously thought. 

Gamma-ray bursts were discovered by the Vela spacecraft. Between 1963
and 1973 the United States launched these satellites whose mission
was, among other purposes, to characterize the space environment in
which future detectors would attempt to verify that the Soviet Union
and other nuclear powers were not detonating nuclear weapons in space
to circumvent the newly negotiated Limited Test-Ban
Treaty.\cite{kle98} The Vela program consisted of pairs of satellites
of progressively greater complexity.  In 1969 R.~Klebesadel noticed a
burst of radiation had been detected by both Vela~4 satellites on July
2, 1967, verifying that this event was external to the detectors;
inspection of the data from subsequent Vela satellites revealed
additional bursts. Since these events did not have the signature of
nuclear explosions, the existence of the phenomenon was never
classified. However bursts were not reported to the astrophysical
community until 1973~\cite{kle73} after later Vela satellites
demonstrated that the sun was not the source.\cite{fenp}  A variety of
progressively more sophisticated burst detectors have
flown since, and the analysis techniques have advanced
accordingly. 

I write this review at a time of great change in the study of bursts.
I joined the BATSE instrument team just before BATSE was placed in
orbit, when we ``knew'' that bursts occurred on the surface of local
neutron stars. Within a few months BATSE showed that this was
incorrect,\cite{mee92} and for six years we debated whether bursts
were merely on the outskirts of our galaxy, or at the edge of the
universe. Since the beginning of 1997 the Italian-Dutch X-ray
satellite {\it Beppo-SAX} has localized a number of bursts to small
error boxes (less than an arcminute in radius) within hours, resulting
in the discovery of a few transients in other wavelength bands. As I
describe in greater depth below, the observations of these transients
have shown nearly conclusively that some, and by Occam's Razor
probably all, bursts are cosmological. The wealth of new observational
data, and the resulting understanding of the burst phenomenon, have
led to a serious confrontation of theory and observation.  In light of
these exciting advances in the field, here I provide background
information which will assist the reader in understanding these
developments.  As a member of the spectroscopy group of the BATSE
instrument team, for the past six years I have devoted most of my
research efforts to burst phenomenology, while occasionally
participating directly in the great debates which have raged in the
field; my research interests undoubtedly affect the emphasis of this
review. 

The burst literature is vast, and it is impossible in a brief review
to cite properly all relevant papers.  Frequently a number of
scientists developed the same concept at the same time. Therefore, on
a given point I reference one or two representative papers which will
lead the reader into the appropriate literature.  I beg the indulgence
of my colleagues whose work I have neglected. 
\section{What Are Gamma-Ray Bursts?}
Gamma-ray bursts are transient events which originate beyond the solar
system.  Emission has been observed during the actual burst only above
$\sim 1$~keV; recently lower energy afterglows have been detected.
Bursts constitute a very diverse population, with properties
characterized by large dynamic ranges. Currently two types of bursts
have been identified:  classical bursts and Soft Gamma Repeaters
(SGRs).  Note that SGRs are identified by their galactic coordinates, 
while bursts are named by the date of their occurrence, GRB~yymmdd.

Four repeating SGRs are currently known.  One appears to be in the
Large Magellanic Cloud (LMC) while the other three are towards the
center of our galaxy.  Their spectra can be described as
optically-thin Bremsstrahlung with a temperature of $\sim25-40$~keV,
and are therefore softer (i.e., lower average photon energy) than the
spectra of classical bursts (as will be discussed below).\cite{nor91}
The most extraordinary burst from an SGR was the first observed event
from SGR~0525-66, which occurred on March 5, 1979;~\cite{cli82} this
burst began with an intense spike and concluded with 25 cycles of an 8
second period. This SGR has been localized to the supernova remnant
N49 in the LMC.\cite{rot94} Similarly SGR~1806-20 is coincident with a
supernova remnant in our Galactic plane.\cite{mur94} On the other
hand, SGR~1900+14 is outside of a supernova remnant; there are no
obvious supernova remnants in the error box of the newly discovered
SGR~1815-14.\cite{hur97} Based on this small sample, an evolutionary
scenario has been suggested in which SGRs are high velocity neutron
stars which eventually escape the supernova remnant created by the
supernova in which the neutron star was born.\cite{kul98} While SGRs
are interesting in their own right, the rest of this review will focus
on classical bursts. 

Classical bursts are known to range in duration from 4~ms to more than
1000~s; the duration distribution is bimodal, with a cusp at a
duration of 2~s.\cite{kou93} The broad spectrum usually peaks between
a few 100 keV and a few MeV; generally most of the energy is emitted
around an MeV, with only a few percent in the X-ray band, although
some events are rich in X-ray emission.  As will be discussed below,
the existence of line features in burst spectra is controversial.  The
light curves vary in appearance:  some bursts are very smooth, while
some are very spiky. Only one class has been defined:  about 15\% of
all bursts are FREDs---Fast Rise, Exponential Decay.  Integrating over
the spectrum, the peak photon flux ranges between 0.1 and 100 photons
cm$^{-2}$ s$^{-1}$, and the energy fluence has been observed between
$10^{-7}$ and $10^{-3}$ erg cm$^{-2}$.  Of course, a quantity's
observed distribution is a convolution of the true distribution and a
given detector's characteristics.  For example, BATSE (the largest
detector system flown thus far) detects bursts by an increase in the
counts accumulated in 64, 256 and 1024~ms bins; consequently BATSE
becomes progressively more insensitive as the burst duration decreases
below 64~ms.\cite{lee96}  Nonetheless, it is clear that burst
properties are characterized by large dynamic ranges. 

Bursts are interesting for many reasons.  Of course, their unknown
origin challenges and entices us to solve the mystery.  Furthermore,
the phenomenon must involve exotic physics.  Somehow gamma-rays are
emitted efficiently with little low energy radiation, at least during
the burst.  The rapid time variability (on scales of less than a
millisecond) indicates a small source size (tens of kilometers) yet
the large distance to the source requires a large energy release (more
than $10^{51}$ ergs if radiated isotropically at cosmological
distances); consequently, the energy density must be enormous.
Finally, if bursts originate in distant galaxies, as appears to be the
case, then they may probe the evolution of the universe. Burst
repetitions and temporal structure within bursts may reveal
gravitational lensing by intervening structure.\cite{nem93} Bursts may
trace the star formation history of galaxies soon after they formed. 
Thus cosmological bursts are intimately tied to the evolution of
matter in the universe. Therefore, the allure of gamma-ray bursts
should not be diminished by the conclusion that they are cosmological
as opposed to Galactic. 
\section{The Mystery Of Burst Origin} 
Before the launch of {\it CGRO} with its BATSE detectors in April,
1991, bursts were thought to originate on local (closer than 200~pc)
neutron stars: bursts were explained as magnetospheric activity on
such stars, and the absorption lines reported between 15 and 100~keV
implied $10^{12}$ gauss fields, comparable to the fields observed on
pulsars (spinning neutron stars). This hypothesis was built on a
variety of arguments which were persuasive but hardly definitive. 

The primary observations regarding the gamma-ray burst distance scale
has been their spatial and intensity distributions; until recently
there had been no definitive signatures in the observations of
individual bursts. Because the observed intensity $f$ is proportional
to the inverse square of the distance $d$ to the burst, $f\propto
d^{-2}$, and the volume out to this distance is proportional to the
cube of the distance, we expect the number of bursts with an intensity
greater than $f$ to be $N(>f)\propto f^{-3/2}$ if the burst sources
are distributed uniformly in three-dimensional Euclidean space.  The
argument holds for various types of bursts with different intrinsic
intensities since each population contributes a power law distribution
$N(>f)\propto f^{-3/2}$.  Thus as long as the source distribution is
uniform, the cumulative intensity distribution will be a -3/2 power
law. Note that this argument applies to any intensity measure which
varies as $d^{-2}$; thus $f$ can be the energy or photon fluence
(total flux integrated over time) or the maximum energy or photon flux
over any particular energy band. Burst detectors trigger on the number
of photons detected in a given energy band over one or more
accumulation timescales.  Therefore the intensity measure most closely
related to the detection process is the peak photon flux (ph s$^{-1}$
cm$^{-2}$) in this energy band. 

The spatial distribution of bursts on the sky reveals the geometry of
the source population.  For example, a Galactic population is expected
to favor the Galactic plane or center when bursts can be seen to
distances of more than a few 100~pc, a typical scale height (i.e., the
distance over which the density of a given constituent of the Galactic
disk decreases perpendicular to the disk).  Because it is very
difficult to focus gamma-rays into images, other methods must be used
to locate bursts. BATSE localizes bursts by comparing the rates in
detectors with different orientations.  The uncertainty of the
resulting localization is typically 5$^\circ$ or more ($\sim2^\circ$
systematic error and a statistical uncertainty which decreases as the
burst intensity increases~\cite{bri98}), which is nonetheless
sufficient to determine whether the bursts are isotropic or favor the
Galactic plane or center.  Strong bursts can be localized to arcminute
uncertainties by comparing the arrival times of the burst signal at
detectors spread throughout the solar system;~\cite{att87} thus far
three interplanetary networks (IPNs) have operated over the past 25
years.  Two detectors localize bursts to an annulus, three detectors
to two points mirrored through the plane of the detectors, and four
detectors can not only localize the burst to a point but can also set
lower limits on the burst's distance. Since bursts were expected to be
a Galactic phenomenon, the spatial distribution is typically
quantified by moments (primarily dipole and quadrupole) in Galactic
coordinates (although other coordinate systems, and coordinate-free
moments have been considered).\cite{bri96a} 

Before BATSE, bursts were observed to be distributed isotropically,
and the intensity distribution was the -3/2 power law expected for a
homogeneous source population.  This was consistent with the
hypothesis that bursts originate on neutron stars in our immediate
vicinity; according to this hypothesis detectors before BATSE were
detecting bursts only out to distances less than the neutron star
population's scale height. Balloon flights with prototype BATSE
detectors showed that BATSE would find a cumulative intensity
distribution flatter than the -3/2 power law at the faint
end.\cite{mee85} The prediction was that the faint (and therefore
distant) bursts would occur preferentially either in the Galactic
plane or towards the Galactic center.  This is analogous to observing
an isotropic sprinkling of stars in the bright night sky above a city,
and discovering the Milky Way in the dark countryside sky. What did
BATSE actually observe? 

The cumulative intensity distribution of the BATSE bursts can be
approximated by two power laws, one with an index -3/2 at the bright
end, and the other with an index of -0.8 at the faint end (see the
left hand side of Figure~1); BATSE has definitely seen beyond the
region where bursts are distributed uniformly.  
\begin{figure}% fig 1
\centerline{\psfig{file=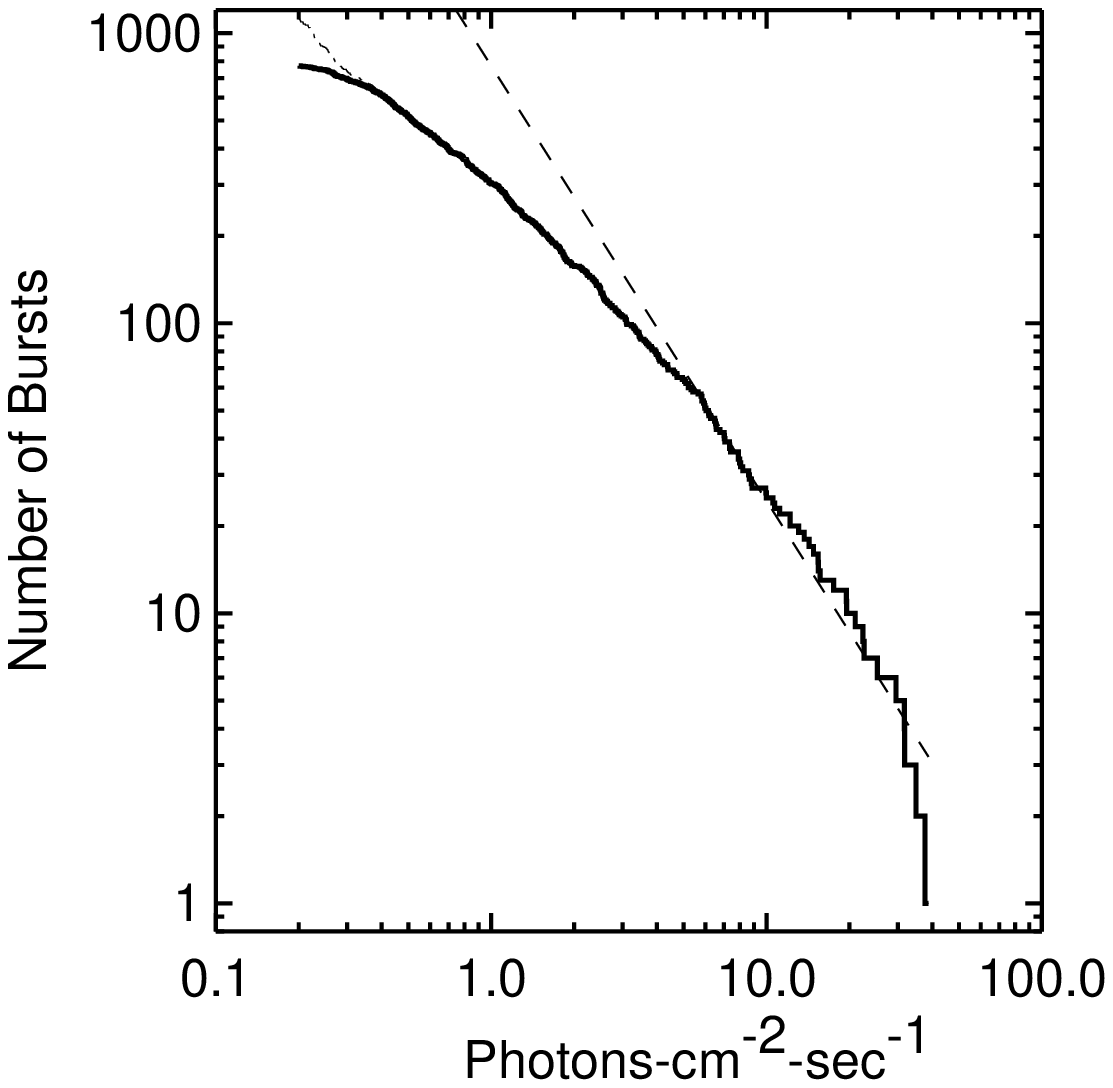,width=5cm}
\psfig{file=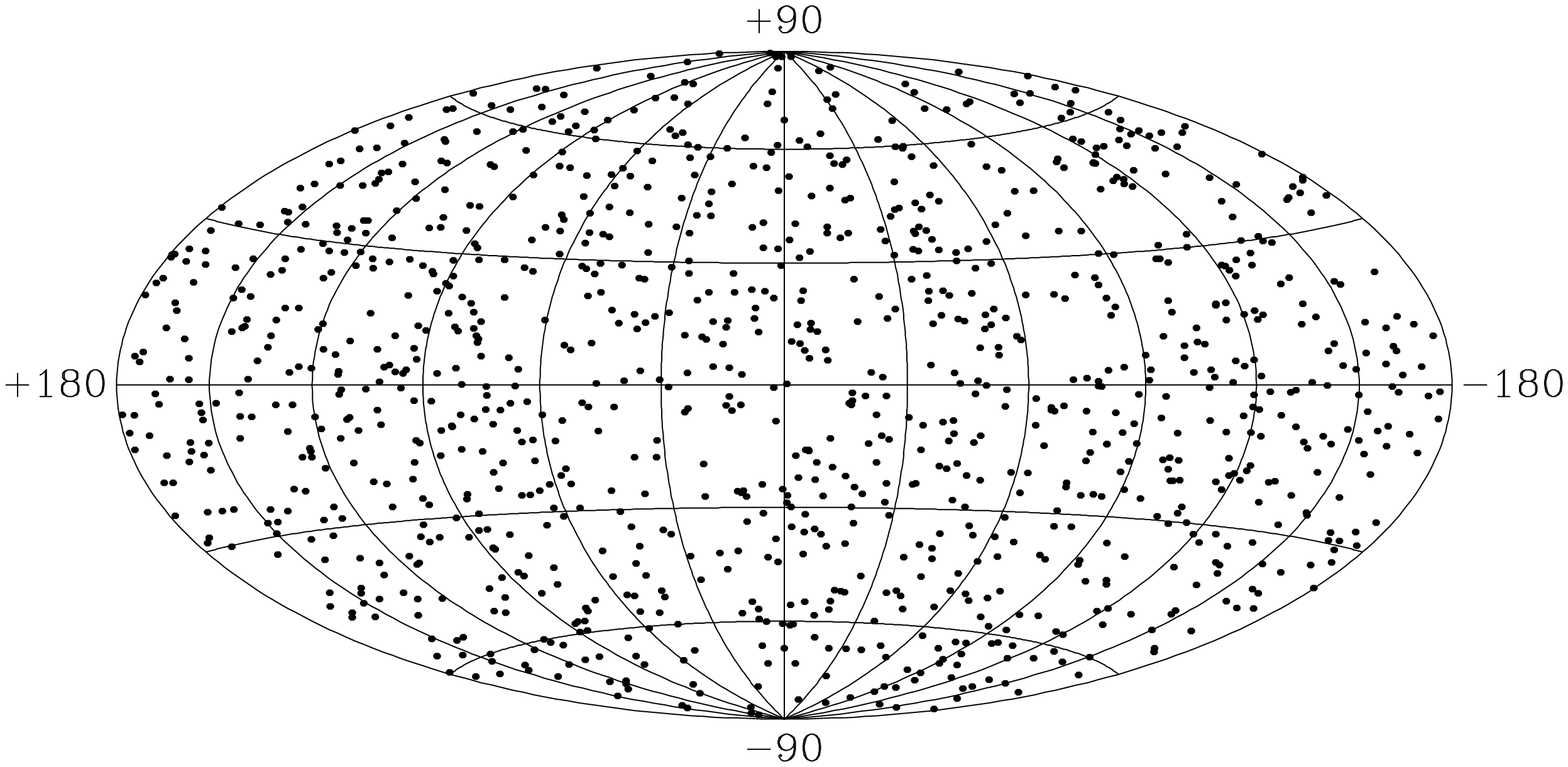,angle=-90,width=7cm}}
\caption{
Cumulative intensity distribution (left) and spatial distribution
(right) of the bursts of the 3rd BATSE Catalog.  The intensity is the
peak photon flux accumulated over 1024~ms in the 50-300~keV energy
band.  The dashed line on the intensity distribution is a -3/2 power
law with a normalization chosen to coincide with the bright portion of
the data.  The dot-dashed curve below 0.3~photon cm$^{-2}$ s$^{-1}$
corrects for the variable detector threshold.  The spatial
distribution is shown in Galactic coordinates.\hfill\null} 
\end{figure}
However, contradicting the hypothesis that bursts originate in the
Galactic plane, the spatial distribution is still consistent with
isotropy (see the right hand side of
Figure~1)!~\cite{mee92,bri96a,mee96}  We are at the center of
a bounded spherical source population. 

Three explanations were advanced based on possible spheres centered on
the earth.  A very few scientists suggested bursts are a solar system
phenomenon, perhaps occurring in the Oort Cloud.\cite{whi94} However,
some preference for the orbital plane is expected,\cite{cla} and no
convincing mechanism was ever advanced. 

A minority of those studying bursts proposed a population of sources
in the Galactic halo.\cite{lam95} The scale of this halo distribution
would have to be large enough to make our offset of 8.5~kpc from the
Galactic center unobservable (1~kpc$=3.1\times10^{21}$~cm). Assuming
the sources emit isotropically, the halo population cannot be too
large or we would observe bursts from nearby galaxies (e.g., the
Andromeda Galaxy, M31) which presumably also are surrounded by burst
sources.  While this hypothesis keeps bursts in the Galaxy, the
distance scale is a thousand times greater than for the local disk
hypothesis, the energy requirement has increased by a factor of a
million, and all the pre-BATSE theories were essentially invalidated. 

Finally, the majority of those who were willing to commit themselves
placed bursts at cosmological distances.\cite{pac95}  Since the
universe is isotropic in the standard cosmology, this explanation
automatically results in an isotropic burst distribution.  The
curvature of space very naturally produces the apparent decrease of
burst sources at large distances without invoking an evolving source
population (although the population undoubtedly does evolve).  The
distance scale is now 10 million times greater than for the pre-BATSE
theories; the energy is therefore 100 trillion times greater. The
required energy is of order $10^{51}$~ergs, or somewhat greater, which
is about the binding energy of a neutron star.  The favored scenario
was the merger of two neutron stars,\cite{eic89,nar92} a cataclysmic
event which destroys the source. 

The burst distribution was shown to be isotropic yet homogenous within
about half a year of BATSE's launch.  Over the subsequent six years
various controversies raged which were surrogates for the debate over
the bursts' distance scale.  For example, Wang and
Lingenfelter~\cite{wan93,wan95} found that five bursts appeared to be
clustered in time and space, while Quashnock and Lamb~\cite{qua93}
found an excess of bursts with small spatial separations; these two
analyses suggested that bursts repeat. Because the amount of energy
required for a cosmological origin almost necessitates a cataclysmic
event which destroys the source, it is highly unlikely that
cosmological sources would repeat.  Also, since the sources are
probably in galaxies, it seems implausible that only a small number of
sources would be active at any time. Subsequent analysis disputed the
significance of the observational evidence for
repeaters;~\cite{nar93,ben96} an improvement of BATSE's burst
localization algorithm revised the burst positions and eliminated
the apparent repetition signal.\cite{mee96} 

If bursts are cosmological then their spectra should be redshifted and
their lightcurves should undergo time dilation.  Faint bursts are
presumably further, and therefore should be more affected by these
relativistic effects.  The difficulty is that burst properties
generally vary by orders of magnitude whereas the cosmological
signatures are factor of 2 or 3 effects.  In addition, intrinsic
correlations between burst properties could mimic the cosmological
signatures, and at most the observations can be shown to be consistent
with the cosmological effect.  That faint bursts have softer (lower
average photon energy) spectra, consistent with a cosmological
redshift, is uncontroversial.\cite{mal95}  The debate has raged over
the presence of time dilation, with small and improperly defined
samples plaguing the analysis by both those who find an effect and
those who do not. Initially Norris \etal~\cite{nor94} found a strong
time dilation signature; Mitrofanov \etal~\cite{mit96} reported that
this signature was absent in their analysis. Fenimore and
Bloom~\cite{fen95b} showed that the apparent time dilation of bursts
at a given $z$ is diminished by spectral redshifting:  temporal
structure is ``narrower'' at high energy (i.e., spikes last longer at
low energy as a result of spectral evolution), and the observed time
dilation is reduced when this narrower structure is redshifted into
the observed energy band. Using a variety of techniques, Norris and
his colleagues continue to observe time dilation,\cite{nor96} although
the effect is smaller than their initial report.  Using a larger
sample than before, Mitrofanov and his colleagues also find time
dilation.\cite{mit98}  It is currently not clear whether all the
studies which find apparent time dilation are consistent. 

A great deal of discussion focused on observations which could solve
the mystery by testing the predictions of the various hypotheses. If
bursts indeed occur outside of the Galactic plane where most of the
absorbing gas is found (the K-shells of ``metals'' such as oxygen in
the interstellar medium absorb X-rays below 1~keV), then bursts' X-ray
spectra should have a low energy cutoff (assuming the intrinsic
spectrum can be estimated).  If bursts arise in large Galactic halos,
then detectors about an order of magnitude more sensitive than BATSE
should detect an excess towards nearby galaxies such as
Andromeda.\cite{bul96} However, the greatest hope was placed in
linking bursts with known astrophysical phenomena, primarily by
finding a counterpart in another wavelength band.  To that end,
systems were developed to monitor the sky continuously (e.g., the
Explosive Transient Camera---ETC~\cite{kri96}---on Kitt Peak) or to
respond rapidly to a burst (e.g., the Gamma-Ray Optical Counterpart
Search Experiment---GROCSE~\cite{lee97,par97a}---at Lawrence Livermore
National Laboratory). These various projects filled in the
three-dimensional space~\cite{mcn95} of:  1) the time since the burst;
2) the wavelength searched; and 3) the depth of the search. For many
years only upper limits were reported.  Great hope was placed in the
{\it High Energy Transient Explorer (HETE)}~\cite{ric92} which had
coaligned gamma-ray, X-ray and ultraviolet detectors, the last two
with spatial resolution. However, the launch vehicle failed to release
{\it HETE}, and the mission was lost; the mission is being rebuilt
with a soft X-ray detector replacing the ultraviolet camera. 
\section{The Solution?} 
The Italian-Dutch X-ray satellite {\it Beppo-SAX} has linked a number
of gamma-ray bursts to counterparts in other wavelength bands, and the
mystery of the burst distance scale may have been solved.  This
satellite includes two wide-field cameras (WFCs) perpendicular to the
coaligned narrow-field telescopes which are the mission's main
instruments. These WFCs each have a field-of-view of 40$^\circ$;
consequently they will observe approximately 8 bursts per year. 
Because of this burst capability, the anti-coincidence shields
surrounding the narrow-field instruments are also used as a burst
detector.  This array of instruments allows {\it Beppo-SAX} to
localize gamma-ray bursts, and search the region where the burst
occurred for an X-ray afterglow. Within $\sim$4--8 hours of the burst
the coordinates can be disseminated to observers in other wavelength
bands, who can then search for the burst source.

The precise sequence of events is as follows:  The burst detector
detects a burst.  When the telemetry from the time of the burst
reaches the ground, images are constructed for the two WFCs using the
photons detected while the burst was in progress (this maximizes the
signal-to-noise ratio).  If the burst was in the field-of-view of one
of the WFCs, one of the images will contain an X-ray point source not
present before the burst.  The source in the WFC image can be
localized to an error radius of 3~arcmin.  The spacecraft is then
reoriented (on a timescale of $\sim$4--8~hours) so that the
narrow-field instruments point at the location of the burst, and one
or more X-ray sources may be found in the burst error box.  Repeated
observations over the next few days may identify a variable source
which is likely to be the burst's afterglow.  The narrow-field
instruments can localize a source to an error radius of
$\sim50$~arcsec.  Optical telescopes can then image these small error
boxes over the next few days, searching for a variable source
(variability is the expected signature of the burst counterpart). 

These error boxes are unprecedented in the study of gamma-ray bursts.
The WFC error box is comparable to the better burst error boxes
resulting from the IPNs, and the position of the X-ray afterglow is
comparable to the very best error boxes previously available.  And the
{\it Beppo-SAX} positions are available within hours of the burst, allowing
the search for fading afterglows.  The scientific bonanza resulting
from the {\it Beppo-SAX} observations has led to other methods of rapidly
localizing bursts. About one burst a month falls within the field of
the All-Sky Monitor on the {\it Rossi X-ray Timing Explorer (RXTE)};
{\it RXTE} raster-scans the resulting error box with its main
detectors. Similarly, {\it RXTE} raster-scans small BATSE error boxes
at the rate of $\sim$once a month.  Thus many afterglows, or limits on
their presence, should be available over the next few years. 

A number of bursts have been localized by either {\it Beppo-SAX} or {\it
RXTE}.  In some cases no afterglow was observed, in others only an
X-ray variable was identified, but in two cases optical transients
were found.  These two optical transients have provided a wealth of
data. 

GRB~970228 was the first burst localized rapidly.  {\it Beppo-SAX}'s
narrow-field instruments,\cite{cos97} and subsequently {\it
ASCA}~\cite{yos97} and {\it ROSAT},\cite{fro97} identified and tracked
a fading X-ray transient. Optical observations 20~hours and 8~days
after the burst found a fading optical source;~\cite{van97} this source
was then observed by a large number of telescopes, including
Keck~\cite{met97b} and {\it Hubble Space Telescope (HST)}.\cite{sah97}
Both the X-ray and optical emission faded at a rate proportional to
$t^{-1.1}$, which can be explained by current theoretical models
(discussed below).\cite{wij97a}  Possibly the host galaxy, extended 
emission underlies the point source.\cite{fru97} Thus this burst
conforms to the expectations for a cosmological source:  a
fading afterglow at the location of the burst superimposed on a
galaxy. Reports that the point source exhibited proper motion
(i.e., moved)~\cite{car97} and that the extended source was
fading,\cite{met97c} which would have contradicted these expectations,
were not verified by a {\it HST} observation in September, 1997, which
found that the transient had not moved, and that the extended source's
brightness was consistent with the previous observations.\cite{fru97} 

GRB~970508 appears to have provided the ``smoking gun'' that at least
some bursts are cosmological.  Once again a fading X-ray transient was
observed.\cite{cos97b}  However, for this burst the variable optical
source brightened for 2 days before it began to fade.\cite{ped97}
Subsequently a radio source coincident with the X-ray and optical
transient was also discovered.\cite{fra97}  Scintillation by our
galaxy's interstellar medium can explain the initial rapid variations
of the radio flux which subsequently damped out; this implies that the
radio source expanded from an initial apparent size of $\sim
10^{17}$~cm.\cite{wax97} Most significantly, absorption lines of
Mg~II and Fe~II at a redshift of $z=0.835$ were found in a spectrum of
the optical transient; a Mg~II system at $z=0.767$ is also
present.\cite{met97} These lines consist of doublets and the line
identifications are therefore extremely secure. Thus the source must
be in or behind this absorption system.  The absence of absorption by
the Ly$\alpha$ forest indicates the source must be at a redshift less
than $z=2.1$. When the optical source faded further, an O[II] emission
line at $z=0.835$ became apparent.\cite{met97d}  No extended source
underlying the transient has been observed,\cite{pia98} although there
are two galaxies $\sim5$~arcsec from the optical source.  The
redshifts of these galaxies have not yet been determined, but 5~arcsec
at a redshift of $z=0.835$ corresponds to $\sim$30~kpc.\cite{sahu97a}
Various possibilities are possible:  the burst progenitor at $z=0.835$
was either in a faint, thus far undetected, galaxy or in the outer
part of a halo of one of the two observed galaxies; or the
line-of-sight to the source which was at $z>0.835$ passed through a
faint galaxy or the halo of one of the observed galaxies at $z=0.835$.

The significant conclusion drawn from the absorption lines is that at
least some bursts are cosmological.  Occam's Razor dictates that
unless proven otherwise, we should assume that all bursts have the
same origin and are thus cosmological.  However, Loredo and
Wasserman~\cite{lor97} have shown that the data permit at least
two burst populations, one of which could be cosmological and the
other local. 
\section{Theories for Cosmological Bursts}
Theories for bursts at cosmological distances developed after the
BATSE observations invalidated the local Galactic neutron star
hypothesis. Because the current theories are discussed in great detail
elsewhere, particularly in these proceedings, here I provide only a
schematic outline. 

The burst originates with the release of a large amount of energy in a
small volume.  The resulting processes erase most of the memory of the
origin of this energy.  The necessary energy release of more than
$10^{51}$ ergs (if bursts radiate isotropically) suggests a source
related to the binding energy of a stellar mass such as the merger of
neutron star-neutron star binaries~\cite{eic89,nar92} or unusual,
extremely energetic supernovae (``hypernovae'').\cite{pac97a} The
necessary rate is approximately once per $10^5$ years per $L_*$ galaxy
(again assuming bursts radiate isotropically).\cite{nar92} 

Neutron star-neutron star binaries are observed to exist and decay by
gravitational radiation, and the rate per galaxy should be
sufficient.\cite{nar91}  Since a comprehensive calculation has not yet
been feasible, it is unknown whether a neutron star-neutron star
merger will release sufficient energy for an observable burst. Davies
\etal~\cite{dav94} used a Newtonian smoothed particle hydrodynamics
code to model the merger.  The merged object is close to the maximum
mass for a spinning neutron star, a disk of material is left in the
equatorial plane, and of order $10^{53}$~ergs is released in various
forms which can be used to power the burst.  Using a
piecewise-parabolic hydrodynamics code, Ruffert and
colleagues~\cite{jan96,ruf97,ruf98} perform Newtonian calculations of a
merger which include gravitational radiation and its back-reaction;
they find that insufficient energy is released. Mathews and
Wilson~\cite{mat97,mat98} calculate the fully relativistic inspiral of
the binary; their numerical methodology does not allow them to follow
the binary to the final merger.  However, they find that as a
consequence of general relativistic effects, the two neutron stars
collapse to black holes before the merger.  But before the collapse,
the neutron stars heat up and radiate a large neutrino flux
($\sim10^{53}$ ergs) before they collapse.  These various calculations
include different physical processes, and thus reach divergent
conclusions. 

Hypernovae are currently only a theoretical construct, and can be
postulated to occur sufficiently frequently. 
Paczy\'nski~\cite{pac97a} proposed a model where a massive rotating
star collapses to a black hole, leaving behind a disk of material
which then accretes onto the black hole, releasing energy.  Fuller and
Shi~\cite{ful97} suggest that the supernova of a supermassive star
($M>5\times 10^4M_\odot$) may emit a large enough neutrino flux to
power the burst. 

The simplest models assume that binary mergers and hypernovae occur in
galaxies and are endpoints of stellar evolution, and therefore a
reasonable conclusion is that bursts occur in galaxies. Of course,
alternatives are possible.  A neutron star-neutron star binary may be
ejected from the galaxy and may not burst until it has traveled a fair
distance from the host galaxy.\cite{blo97}  The supermassive star
($M>5\times 10^4M_\odot$) which might power the burst may form
outside of a galaxy.\cite{ful97} 

If the gamma-ray energy density is sufficiently large, the resulting
volume will be optically thick to pair creation.  A pair plasma should
result which will expand relativistically; the Lorentz factor $\Gamma$
of the fireball depends on the ratio of the energy to the number of
baryons which are swept up by the plasma.  The original fireball
models~\cite{goo86,pac86} attributed the gamma-ray emission to the
moment when the fireball becomes optically thin. However, this should
produce a single short spike with a quasi-black body spectrum, and is
therefore insufficient to reproduce the observed spectra and temporal
structures.  In the next generation of models the ``external'' shocks
which form when the fireball collides with the surrounding medium
radiate the observed gamma-rays.\cite{mes94}  However, the external
shocks are not thought to be capable of producing the rich temporal
structure unless the shocks radiate with very low
efficiency.\cite{sar97} Consequently, in the current fireball model
inhomogeneities within the relativistic outflow result in ``internal''
shocks which radiate the observed gamma-ray
emission.\cite{nar92,ree94,pap96}  However, the ``external'' shocks
should radiate at lower energies on timescales much longer than the
gamma-ray burst; this is the origin of the recently observed
afterglows.\cite{mes97,wij97a} Here I have provided only a very brief
outline of the fireball model, and I encourage the reader to consult
the vast literature (including the papers in these proceedings). 
\section{The ``Minimal'' Cosmological Scenario}
The burst intensity distribution is the convolution of the intrinsic
luminosity function and the burst rate, both as functions of the
distance to the burst source.  The simplest (i.e., ``minimal'') model
assumes that there has been no cosmological evolution in the burst
rate per comoving volume or in the luminosity function.  Further, the
luminosity function is assumed to be a delta function for a certain
intrinsic intensity measure (e.g., the total radiated energy or the
peak photon luminosity), that is, all bursts are ``standard candles.''
The standard candle has been the same at all cosmological epochs by
the no-evolution assumption.  The aesthetic beauty of the cosmological
scenario is that the bend in the intensity distribution can be wholly
explained by the curvature of space and time dilation at cosmological
distances. A consequence of the minimal model is that there is a
one-to-one mapping between the burst intensity and the distance, with
the distance scale given by the shape of the intensity distribution.
For example, by this model the faint BATSE bursts are at a redshift of
$z\sim1$.\cite{fen93} 

Which intensity is the standard candle is of course uncertain. 
Because BATSE triggers on the count rate in the 50-300~keV band, and
the BATSE bursts constitute the largest homogeneous database, the peak
photon flux (corresponding to the peak photon luminosity) is often
used.  An intensity measure related to a detector's trigger is favored
because the low intensity threshold is best understood.  However, this
is a choice based on instrumental considerations and not on physics. 

Many of the proposed energy sources for cosmological bursts are the
endpoints of stellar evolution, and consequently a simple assumption
is that bursts occur in galaxies at a rate proportional to a galaxy's
mass.\cite{fen93}  Assuming a constant mass-to-light ratio, this
implies that the rate is proportional to a galaxy's luminosity. 
Therefore, the minimal theory predicts that the host galaxy's
luminosity function is the luminosity function for regular galaxies,
weighted by one power of the luminosity. 

The minimal model is based on unrealistically simplistic assumptions.
The great variety of burst profiles and the large dynamic range of
burst properties such as spectral hardness and time duration make the
standard candle assumption suspect.  Indeed, Hakkila \etal~\cite{hak96}
found that a standard candle cosmological model, where the peak energy
flux (ergs cm$^{-2}$ s$^{-1}$) corresponds to the standard candle,
does not fit the joint PVO-BATSE distribution (of course, a different
intensity measure might correspond to the standard candle). Similarly,
all known astrophysical phenomena have undergone cosmological
evolution. Studies have constrained the cosmological evolution of the
burst rate.\cite{fen95b}  While it has generally been recognized that
the burst population must have undergone evolution, and that bursts at
any given epoch were not standard candles, most studies fitting the
burst database have adopted the minimal scenario. 
\section{Complications with the ``Minimal'' Cosmological Scenario}
As described above, the minimal cosmological scenario predicts that
bursts should occur in galaxies, and that the distance to the burst,
and therefore to the galaxy, can be calculated from the intensity.
Schaefer~\cite{sch92} pointed out that the small error boxes of 8
bright bursts do not contain bright galaxies; if the brightest galaxy
in the error box, or the detection threshold for the box, had a
brightness equal to M31 (the Andromeda Galaxy), the total burst energy
must have been as large as $2\times10^{53}$ ergs. Fenimore
\etal~\cite{fen93} found that Schaefer's data were only marginally
consistent with the galaxies predicted by the minimal scenario if the
brightest galaxy in each error box was indeed the host galaxy. 
However, the brightest galaxy could also be an unrelated background
galaxy.  This apparent discrepancy with the minimal scenario has been
dubbed the ``no host galaxy'' problem.  On the other hand, Larson and
collaborators~\cite{lar96,lar97a,lar97b} reported that their sample of
error boxes, which were somewhat larger than Schaefer's, had an {\it
excess} of bright galaxies, although they recognized that they could
not distinguish between host and background galaxies. 

D.~Hartmann and I realized that a more sophisticated analysis
methodology was required.\cite{ban98} Therefore we use a likelihood
ratio which contrasts the hypothesis that both host and unrelated
background galaxies are present with the hypothesis that all the
observed galaxies are unrelated background galaxies. This ratio was
developed within a Bayesian framework, but it is understandable within
standard ``frequentist'' statistics. If this ratio is much greater
than 1 then a host galaxy is clearly present in each error box, while
if the ratio is much less than 1 then no host galaxy is present.
Finally, if the ratio is of order unity then the data are
inconclusive.  By construction, this methodology accounts for the
unrelated background galaxies which will be detected if the error box
is searched deeply enough.  We include each detected galaxy in
addition to the detection limit, and we permit a more sophisticated
description of the error box.  This methodology demonstrates that the
observations of a given error box can show conclusively that the host
galaxy is present only if the expected host galaxy is on average
brighter than the average brightest background galaxy, which depends
on the size of the error box. 

Thus far we have applied this methodology to only a few datasets, but
the results show that the minimal scenario is indeed too simple.  We
find that the likelihood ratio for the nine fields observed by Larson
and McLean~\cite{lar97a} is 0.25, which indicates that we are unable to
determine whether host galaxies are present.  On the other hand, the
likelihood ratio for the four error boxes observed by Schaefer
\etal~\cite{sch97} with the {\it HST} is $2\times 10^{-6}$ which
clearly shows that the host galaxies predicted by the minimal scenario
are not present. 

One would think that the optical transients resulting from the {\it
Beppo-SAX} observations would produce more conclusive results than the
larger error boxes previously available.  The optical transient left
by GRB~970228 sits on a region of extended emission with a flux of
$V=25.7$;~\cite{fru97} this extended emission appears to be the host
galaxy we expected!  However, when we apply our methodology, we
calculate a likelihood ratio of only 0.27, {\bf not} a number much
greater than one. The reason for such a small value is that a galaxy
with $V=25.7$ is typically at a redshift of $z\sim2$, not the $z\sim
0.25$ calculated from the burst's intensity.  There is no hint of
extended emission underlying the GRB~970508 transient down to a
magnitude of $R\sim25.5$.\cite{pia98}  Assuming the burst was at
$z=0.835$ (this is actually the lower limit, but the burst intensity
was consistent with this redshift), the likelihood ratio is 0.027.
Thus the two recent well-localized bursts also show that the host
galaxies predicted by the minimal scenario are not present. 

Consequently alternative scenarios have been suggested.  The burst
rate might be proportional to the star formation rate; for example, a
hypernova may result from a short-lived massive star, and thus bursts
will occur when and where there has been recent star formation. The
universe's star formation history has recently been determined
empirically, and it shows that the rate per comoving volume increased
slowly from $z\sim 5$ to $z\sim 1.2$, and has plummeted
since.\cite{mad96} Using this star formation rate as the burst rate
can reproduce the burst intensity distribution, with the bursts
occurring at much greater distances.\cite{wij97b,tot97} A surprising
consequence is that the portion of the intensity distribution which is
a power law with an index of -3/2 results not from a uniform burst
density in nearby Euclidean space (as discussed in \S 2) but from the
balance between spatial curvature and burst evolution.  It has also
been suggested that at moderate redshifts star formation occurred
preferentially in small galaxies.\cite{sahu97a} Finally, the source
may have been ejected from the host galaxy.\cite{blo97} 

Thus the minimal cosmological scenario is too simple, as was suspected
on astrophysical grounds.  The development of more sophisticated
cosmological theories involves issues such as star formation, and
consequently the study of gamma-ray bursts will be more closely
integrated with cosmology and extragalactic astrophysics. 
\section{Burst Phenomenology} 
The discovery of the likely burst distance scale, and the additional
information provided by the burst afterglows, have motivated more
detailed theories.  Consequently, burst phenomena which were
relatively unimportant for determining the distance scale have become
important for revealing the physics of the burst process. 
\subsection{Spectrum}
The burst continuum from 10~keV to 100~MeV has a very simple shape: it
is curved at low energies and becomes a power law at high energy.
Indeed, the spectrum over these four decades can be characterized by a
four-parameter function~\cite{ban93}
\begin{equation}
N(E) = \cases{
A E^\alpha e^{-E/E_0} & $E < (\alpha - \beta)E_0$ \cr
A^\prime E^\beta & $E \ge (\alpha - \beta)E_0$ \cr
}
\end{equation}
where $A^\prime$ is chosen so that $N(E)$ is continuous and
differentiable at $E = (\alpha - \beta)E_0$.  Figure~2 shows a fit to
a spectrum accumulated over a particularly intense burst. All four
parameters vary within and between bursts, but typically $\alpha\sim
-1$ and $\beta\sim -2$.  The energy $E_p$ at which $E^2 N(E) \propto
\nu F_\nu$ (the energy flux per energy decade) peaks characterizes
whether a spectrum is hard or soft.  If $\beta < -2$ then
$E_p=(2+\alpha)E_0$, otherwise $E_p$ is above the energy where the
high energy power law rolls over (such a rollover is not included in
eq.~[1]). Usually $E_p$ is calculated from the low energy component,
regardless of the value of $\beta$.  The observed $E_p$ distribution
is between 50~keV and 1~MeV, with a maximum at $\sim$150~keV; see
Figure~3. The true $E_p$ distribution may be broader because the
energy band over which a detector triggers introduces a selection
effect; in particular, there may be a large number of bursts with a
high $E_p$.\cite{pir96}  The $E_p$ distribution is important since the
relativistic fireball models link $E_p$ to the fireball's Lorentz
factor. 
\begin{figure} % fig 2
\centerline{\psfig{file=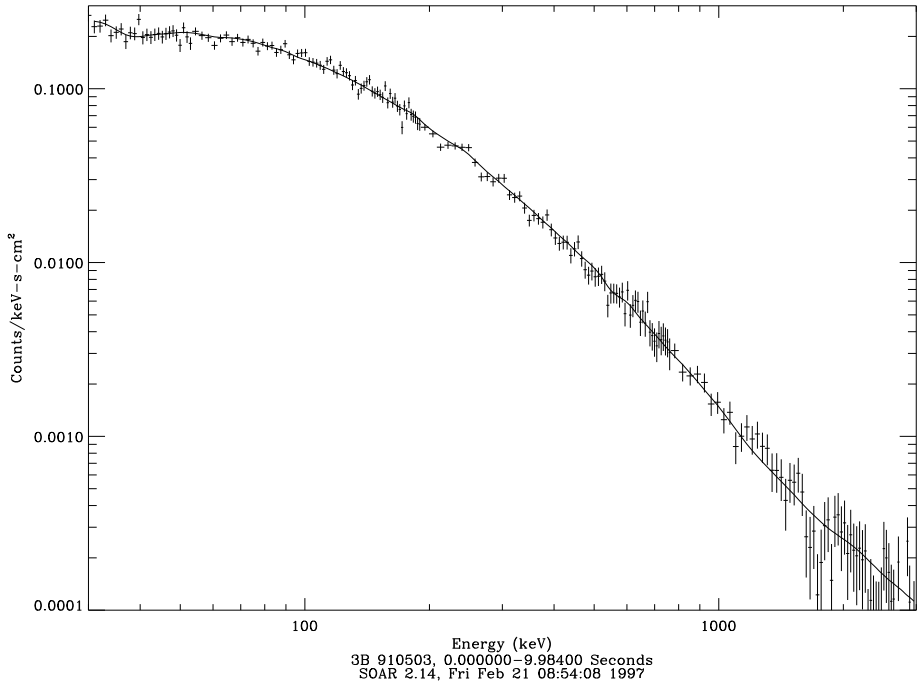,angle=90,width=12.5cm}}
\caption{Fit to the count spectrum accumulated over the first 12~s of
GRB~910503 by a BATSE Spectroscopy Detector.  The solid curve is the
best-fit model folded through the detector response. The detector is
based on a NaI(Tl) crystal, which introduces features into the count
spectrum near the iodine K-edge at 33.17~keV. \hfill\null} 
\end{figure}
\begin{figure} % fig 3
\centerline{\psfig{file=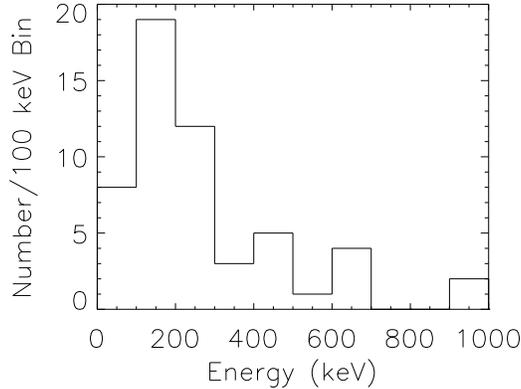,width=8cm}}
\caption{Hardness distribution for a sample of 54 bursts.  The
hardness $E_p$ is the energy of the peak of $E^2N(E) \propto \nu
f_\nu$. \hfill\null} 
\end{figure}

Deviations from this simple functional form are seen at high and low
energies.  Approximately 10\% of a sample of bright BATSE bursts have
low energy excesses; unfortunately this excess was discovered using a
single broad low energy channel, and spectral information on this
excess is unavailable.\cite{pre96}  One of the 22 bursts observed by 
{\it Ginga} also has a low energy excess; otherwise the four parameter 
function of eq.~(1) describes the {\it Ginga} spectra between 2 and 
400~keV.\cite{str98}  The EGRET instrument on {\it CGRO}
detects individual high energy photons with energies above 30~MeV with
a spark chamber.  Usually too few high energy photons are recorded to
provide detailed spectral information.  Nonetheless, there is a
tendency for the emission at a $\sim$GeV to linger after the lower
energy burst.\cite{din94} In GRB~940217, a particularly bright burst,
the high energy emission continued for $\sim$90 minutes after the
160~second burst, with an 18~GeV photon, the highest energy burst
emission yet observed, detected towards the end of this 90 minute
period.\cite{hur94} 
\subsection{Spectral Evolution}
Since the observed gamma-ray burst spectrum reflects the energy
content and particle distributions within the source's emitting
region, spectral variations during a burst are an important diagnostic
of the nature of this region. Early  studies of spectral evolution
reached apparently contradictory conclusions:  Golenetskii
\etal~\cite{gol83} reported that the intensity and spectral hardness
were correlated, while Norris \etal~\cite{nor86} found a hard-to-soft
trend. Subsequent studies using SIGNE~\cite{kar94} and
BATSE~\cite{for95} spectra showed that both trends hold in general: the
spectrum does indeed harden during intensity spikes, but there is a
hard-to-soft trend during these spikes, and the hardness tends to peak
at successively lower values from spike to spike. 

This characterization of spectral evolution resulted from fitting 
a sequence of spectra accumulated during a burst, and comparing 
the time series of a hardness measure such as $E_p$ to the intensity 
lightcurve.  Many counts are required for a good fit to a spectrum, 
and therefore fitting sequences of spectra is feasible only for bright, 
long duration bursts.  Even for the brightest bursts the time necessary 
to accumulate a spectrum with a sufficient signal-to-noise ratio 
(typically more than a second) is usually longer than the time structure 
evident to the eye (the separation between intensity spikes is typically 
a second).  Therefore I have been developing other techniques of 
studying spectral evolution.

To characterize the spectral evolution of a large sample of bursts I
used the auto- and crosscorrelation functions (ACF and CCFs,
respectively) of burst lightcurves in different energy
channels.\cite{ban97b}  BATSE provides discriminator rates in 4 energy
bands (Ch.~1: 25--50, Ch.~2: 50--100, Ch.~3: 100--300, and Ch.~4:
300--2000~keV) on a 64~ms timescale during a burst. I calculated the
CCFs of a fiducial energy channel, Ch.~3 (100--300~keV), with each of
the 4 energy channels (the CCF of the fiducial channel with itself is
that channel's ACF).  By comparing the time lags of the peaks of each
curve and their relative values at different lags, as shown by the
example in Figure~4, I characterized the type of spectral evolution. 
\begin{figure} % fig 4
\centerline{\psfig{file=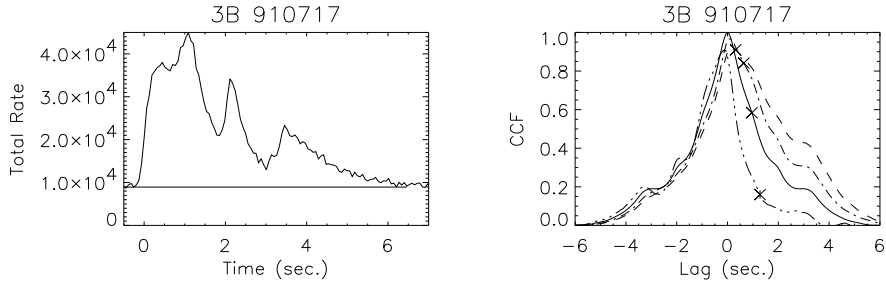,width=12cm}}
\caption{Lightcurve (left) and auto- and crosscorrelations for the 4
energy channels (right) for GRB~910717. The order of the correlation
curves on the positive lag side indicates hard-to-soft spectral
evolution.\hfill\null} 
\end{figure} 

I calculated the ACFs and CCFs for 209 strong, mostly long
bursts.\cite{ban97b}  The order of the CCF peaks shows that in general
high energy emission precedes low energy emission. As was known
previously from comparing the ACFs of the different
channels,\cite{fen95} the CCF widths indicate that high energy
temporal structure is narrower than low energy structure (i.e., spikes
last longer at low energy than at high).  The relative order of the
CCFs at different lags shows there is hard-to-soft evolution within
and among spikes in $\sim$80--90\% of the bursts, and there are only a
few cases of soft-to-hard evolution.  The peaks of the CCFs
for the high energy channels typically lead those of the low energy
channels by 0.1-0.2~s.  Thus this study showed that hard-to-soft
spectral evolution is ubiquitous but counterexamples exist. 

Liang and Kargatis~\cite{lia96} found that when the logarithm of $E_p$
is plotted as a function of the cumulative photon fluence (i.e., the
photon fluence from the beginning of the burst to the time $E_p$ is
measured), the datapoints fall on a series of straight lines with the
same slope for a given burst.  This can be explained by an emission
region with a fixed number of radiating particles which is
re-energized for each intensity spike. 
\subsection{Spectral Lines}
Spectral lines provide a great deal of information about the emitting
region.  Missions prior to BATSE---Konus,\cite{maz82}
HEAO-1,\cite{hue87} and {\it Ginga}~\cite{mur88,gra92}---detected
absorption lines between 10 and 100~keV which were attributed to
cyclotron resonant scattering (which scatters photons out of the line
of sight) in $10^{12}$ gauss fields.\cite{ale89,wan89}  Since neutron
stars are the only known anchors for fields of this strength, these
observations supported the hypothesis that bursts originate on local
neutron stars.  However, BATSE has thus far not detected any
lines,\cite{pal94,ban96} and consequently there has been little
interest in explaining spectral lines in a cosmological burst model. A
large and stable (temporally and spatially over the absorption region)
magnetic field is necessary to produce the narrow lines observed by
{\it Ginga}, and creating such a field configuration in a relativistic
fireball will be a major theoretical challenge. 

For the past six years the BATSE spectroscopy team has been searching
for lines and evaluating the results of this search.  The reports of
the absence of a BATSE detection~\cite{pal94,ban96} were based on a
visual inspection of spectra.  A more comprehensive computerized
search has been carried out,\cite{bri96b} and promising line candidates
have been identified which are now being evaluated; we expect to issue
a definitive report in the next year. 

The question of whether the BATSE nondetections are consistent with
the detections by previous missions led me to develop a statistical
methodology to study the consistency between the results of these
missions.\cite{ban94}  For this statistical analysis detailed
information is required not only about the detections but also about
the nondetections; such data are available from BATSE and {\it Ginga}.
This methodology can also extract other physical information, such as
the likely frequency with which lines occur.  The methodology requires
simulations of a detector's ability to detect lines~\cite{ban95}
and models for the occurrence of lines within a burst.\cite{ban97}  
Thus far only preliminary results have been extracted, in part because
only a subset of the necessary {\it Ginga} data has been
processed.\cite{fen93b}  With various approximations, I find that the
two missions are consistent at the few percent level. However, it is 
also clear that lines are not very common (i.e., they may be present 
in only a few percent of all bursts).
\section{Final Word}
An optical transient following a gamma-ray burst is superimposed on an
extended source which might be the host galaxy, and the spectrum of
another transient has absorption lines at $z=0.835$.  Therefore bursts
are cosmological, the mystery has been solved, and the study of this
phenomenon can fade into obscurity as yet another subfield of
astrophysics.  But has the mystery really been solved? 

First, that bursts are cosmological rests on only two optical
transients.  The GRB~970228 transient is coincident with an extended
source which has not yet been proven to be a galaxy and which is
fainter than expected for the host. The $z=0.835$ absorption lines in
the spectrum of the GRB~970508 transient show that the transient is
beyond this redshift, but there is no obvious host galaxy.  Further,
the position of the GRB~970508 X-ray transient is known to only
50~arcsec \cite{cos97b} as opposed to the 10~arcsec uncertainty for
the GRB~970228 X-ray transient,\cite{fro97} and therefore skeptics can
still claim that the optical transient in the GRB~970508 X-ray
transient error box may be unrelated to the burst; the transient sky
has yet to be characterized, particularly at faint optical magnitudes.
In addition, even if these optical transients result from
cosmological bursts, there may yet be a population of Galactic bursts.
I suspect that GRB~970228 and GRB~970508 are indeed cosmological, and
that by Occam's Razor we should assume that all bursts are
cosmological, but we should be aware that this conclusion is still
based on only two bursts. 

Second, even if we interpret the observations as demonstrating that
bursts are cosmological, the analysis of the host galaxy searches
shows that the minimal cosmological model is incorrect (\S 7).  The
alternatives are that bursts do not occur in regular galaxies, or that
they are further than previously thought.  Thus we are uncertain about
bursts' environment and distance scale.  We can hardly claim that the
burst location mystery has been solved. 

Third, the conclusion that the observed emission results from a
relativistic fireball is based on the large energy which is released
in a small volume; few observational signatures of such a fireball
have been identified in bursts' spectral and temporal behavior.
Specifically, we do not know whether the observed gamma-ray radiation
results from synchrotron, inverse Compton or some other emission
mechanism.  The observed spectral evolution is unexplained,
particularly the softening of successive intensity spikes.  Thus the
origin of the observed emission is still a mystery. 

Fourth, even if a relativistic fireball produces the observed
emission, the ultimate energy source is unknown since the fireball
erases almost all memory of its origin.  As I discussed, the merger of
a neutron star-neutron star binary has been proposed as the energy
source, but the admittedly incomplete calculations carried out to date
do not verify the favored scenario.  If bursts originate at higher
redshifts than implied by the minimal cosmological model, then the
energy requirements may exceed the output of the merger of solar mass
scale objects (the angular extent of the gamma-ray emission and
therefore the total energy radiated are unknown).  Consequently, other
energy sources have been suggested, such as the supernovae of massive
stars.  Hence the origin of the bursts' energy is still a mystery. 

The where, how and why of the burst phenomenon remain uncertain.
Further, it is clear that bursts involve extreme physics:  the release
of a large energy in a small volume on a short timescale, resulting in
a relativistic fireball, possibly entraining substantial magnetic
fields. Finally, a deeper understanding of the origin of bursts may
require the history of matter on cosmological timescales; for example
these events may trace the starbursts accompanying galaxy formation. 
Therefore, the study of gamma-ray bursts will remain an exciting and
lively field for the foreseeable future. 
\section*{Acknowledgments}
I thank my colleagues on the BATSE team and at UC San Diego for their
assistance over the past 7 years. My research on gamma-ray bursts is
supported by the {\it CGRO} Guest Investigator Program and by NASA
contract NAS8-36081. 

\input{grbrev97_bib.tex}
\end{document}

%% file: grbrev97_bib.tex
\section*{References}

%% file: grbrev97.bbl
\begin{thebibliography}{99}

\bibitem{kle73}R. Klebesadel, I. Strong, and R. Olson, 
\Journal{\APJL}{182}{L85}{1973}.

\bibitem{fis89}G. J. Fishman, \etal, 
in {\it Proc. of the Gamma-Ray Observatory Science Workshop}, 2-39, 3-47 
(1989).

\bibitem{lam95}D. Q. Lamb, \Journal{\it P. A. S. P.}{107}{1152}{1995}.

\bibitem{pac95}B. Paczy\'nski, 
\Journal{\it P. A. S. P.}{107}{1167}{1995}.

\bibitem{kle98}R. W. Klebesadel, in {\it 4th Gamma-Ray Burst
Symposium}, eds. C. Meegan, R. Preece and T. Koshut (AIP, New York,
1998), in press. 

\bibitem{fenp}E. E. Fenimore, private communication (1997).

\bibitem{mee92}C. A. Meegan, \etal, 
\Journal{\it Nature}{355}{143}{1992}.

\bibitem{nor91}J. Norris, \etal, \Journal{\APJ}{366}{240}{1991}.

\bibitem{cli82}T. L. Cline, \etal, \Journal{\APJL}{255}{L45}{1982}.

\bibitem{rot94}R. Rothschild, S. Kulkarni and R. Lingenfelter,
\Journal{\it Nature}{368}{432}{1994}.

\bibitem{mur94}T. Murakami, \etal, \Journal{\it Nature}{368}{127}{1994}.

\bibitem{hur97}K. Hurley, \etal, IAU Circ. 6743.

\bibitem{kul98}S. R. Kulkarni, in {\it 4th Gamma-Ray Burst
Symposium}, eds. C. Meegan, R. Preece and T. Koshut (AIP, New York,
1998), in press. 

\bibitem{kou93}C. Kouveliotou, \etal, \Journal{APJL}{413}{L101}{1993}.

\bibitem{lee96}T. T. Lee and V. Petrosian, 
\Journal{\APJ}{470}{479}{1996}.

\bibitem{nem93}R. J. Nemiroff, \etal, \Journal{\APJ}{414}{36}{1993}.

\bibitem{bri98}M. S. Briggs, \etal, in {\it 4th Gamma-Ray Burst
Symposium}, eds. C. Meegan, R. Preece and T. Koshut (AIP, New York,
1998), in press. 

\bibitem{att87}J.-L. Atteia, \etal, \Journal{\APJS}{61}{305}{1987}.

\bibitem{mee85}C. A. Meegan, G. J. Fishman and R. B. Wilson,
\Journal{\APJ}{291}{479}{1985}.

\bibitem{bri96a}M. S. Briggs, \etal, \Journal{\APJ}{459}{40}{1996}.

\bibitem{mee96}C. A. Meegan, \etal, 
\Journal{\it Ap. J. Suppl.}{106}{65}{1996}.

\bibitem{whi94}R. S. White, in {\it Gamma-Ray Bursts, AIP Conf. Proc.
307}, eds. G.~J. Fishman, J.~J. Brainerd, and K.~Hurley (AIP, New
York, 1994), 620. 

\bibitem{cla}T. E. Clarke, O. Blaes and S. Tremaine,
\Journal{\it A. J.}{107}{1873}{1994}.

\bibitem{eic89}D. Eichler, M. Livio, T. Piran, and D. Schramm, 
\Journal{\it Nature}{340}{126}{1989}.

\bibitem{nar92}R. Narayan, B. Paczy\'nski, and T. Piran,
\Journal{\it Ap. J. Lett.}{395}{L83}{1992}.

\bibitem{wan93}V. C. Wang and R. E. Lingenfelter, 
\Journal{\APJL}{416}{L13}{1993}.

\bibitem{wan95}V. C. Wang and R. E. Lingenfelter, 
\Journal{\APJ}{441}{747}{1995}.

\bibitem{qua93}J. M. Quashnock and D. Q. Lamb, 
\Journal{\MNRAS}{265}{L59}{1993}.

\bibitem{nar93}R. Narayan and T. Piran, 
\Journal{\MNRAS}{265}{L65}{1993}.

\bibitem{ben96}D. P. Bennet and S. H. Rhie,
\Journal{\APJ}{458}{293}{1996}.

\bibitem{mal95}R. S. Mallozzi, \etal, \Journal{\APJ}{454}{597}{1995}.

\bibitem{nor94}J. P. Norris, \etal, \Journal{\APJ}{424}{540}{1994}.

\bibitem{mit96}I. G. Mitrofanov, \etal, \Journal{\APJ}{459}{570}{1996}.

\bibitem{fen95b}E. E. Fenimore and J. S. Bloom, 
\Journal{\APJ}{453}{25}{1995}.

\bibitem{nor96}J. P. Norris, in {\it
Gamma-Ray Bursts, AIP Conf. Proc. 384}, eds. C.~Kouveliotou, M.~F.
Briggs and G.~J.~Fishman (AIP, New York, 1996), 13. 

\bibitem{mit98}I. G. Mitrofanov, \etal, in {\it 4th Gamma-Ray Burst
Symposium}, eds. C. Meegan, R. Preece and T. Koshut (AIP, New York,
1998), in press. 

\bibitem{bul96}T. Bulik, P. S. Coppi, and D. Q. Lamb, in {\it
Gamma-Ray Bursts, AIP Conf. Proc. 384}, eds. C.~Kouveliotou, M.~F.
Briggs and G.~J.~Fishman (AIP, New York, 1996), 340. 

\bibitem{kri96}H. Krimm, R. Vanderspek and G. Ricker, in {\it
Gamma-Ray Bursts, AIP Conf. Proc. 384}, eds. C.~Kouveliotou, M.~F.
Briggs and G.~J.~Fishman (AIP, New York, 1996), 661. 
 
\bibitem{lee97}B. Lee, \etal,
\Journal{\it Ap. J. Lett.}{482}{L125}{1997}.

\bibitem{par97a}H.-S. Park, \etal,
\Journal{\it Ap. J.}{490}{99}{1997}. 

\bibitem{mcn95}B. J. McNamara, \etal,
\Journal{\it Ap. J. Lett.}{452}{L25}{1995}

\bibitem{ric92}G. Ricker, \etal, in {\it Gamma-Ray Bursts: 
Observations, Analyses and Theories}, eds. C.~Ho, R.~I.~Epstein and
E.~E.~Fenimore (Cambridge Univ. Press, Cambridge, 1992), 288. 

\bibitem{cos97}E. Costa, \etal, \Journal{\it Nature}{387}{783}{1997}.

\bibitem{yos97}A. Yoshida, \etal, IAU Circ. 6593 (1997).

\bibitem{fro97}F. Frontera, \etal, IAU Circ. 6637 (1997).

\bibitem{van97}J. van Paradijs, \etal,
\Journal{\it Nature}{386}{686}{1997}.

\bibitem{met97b}M. R. Metzger, \etal, IAU Circ. 6582 (1997).

\bibitem{sah97}K. C. Sahu, \etal,
\Journal{\it Nature}{387}{476}{1997}.

\bibitem{wij97a}R. A. M. J. Wijers, M. Rees and P. M\'esz\'aros,
\Journal{\MNRAS}{288}{L51}{1997}.

\bibitem{fru97}A. Fruchter, \etal, IAU Circular 6747 (1997).

\bibitem{car97}P. A. Caraveo, \etal, {\it A\&A}, astro-ph/9707163 (1997).

\bibitem{met97c}M. R. Metzger, \etal, IAU Circ. 6631 (1997).

\bibitem{cos97b}E. R. Costa, \etal, IAU Circ. 6649 (1997).

\bibitem{ped97}J. Pedersen, \etal, 
\Journal{\APJ}{496}{astro-ph/9710322}{1997}.

\bibitem{fra97}D. A. Frail and S. R. Kulkarni, IAU Circ. 6662 (1997).

\bibitem{wax97}E. Waxman, S. Kulkarni, and D. Frail, {\it Ap. J. 
Lett.}, astro-ph/9709199 (1997).

\bibitem{met97} M. R. Metzger, \etal,
\Journal{\it Nature}{387}{878}{1997}.

\bibitem{met97d}M. R. Metzger, \etal, IAU Circ. 6676 (1997).

\bibitem{pia98}E. Pian, \etal, {\it Ap. J. Lett.}, 
astro-ph/9710334 (1998).

\bibitem{sahu97a}K. C. Sahu, \etal, \Journal{\APJL}{489}{L127}{1997}.

\bibitem{lor97}T. J. Loredo and I. M. Wasserman, {\it Ap. J. Supp.},
astro-ph/9701112 (1997).

\bibitem{pac97a}B. Paczy\'nski, 
{\it Ap. J. Lett.}, astro-ph/9710086 (1997).

\bibitem{nar91}R. Narayan, T. Piran and A. Shemi, 
\Journal{\APJL}{379}{L17}{1991}.

\bibitem{dav94}M. B. Davies, \etal, \Journal{\APJ}{431}{742}{1994}.

\bibitem{jan96}H. T. Janka and M. Ruffert, 
\Journal{\it A\&A}{307}{L33}{1996}. 

\bibitem{ruf97}M. Ruffert, \etal, \Journal{\it A\&A}{319}{122}{1997}.

\bibitem{ruf98}M. Ruffert, \etal, these proceedings (1998).

\bibitem{mat97}G. J. Mathews and J. R. Wilson,
\Journal{\APJ}{482}{929}{1997}.

\bibitem{mat98}G. J. Mathews, \etal, these proceedings (1998).

\bibitem{ful97}G. M. Fuller and X. Shi, {\it Ap. J. Lett.}, astro-ph/9711020
(1997).

\bibitem{blo97}J. S. Bloom, N. R. Tanvir, and R. A. M. J. Wijers, 
{\it M. N. R. A. S.}, astro-ph/9705098 (1997).

\bibitem{goo86}J. Goodman, \Journal{\it Ap. J. Lett.}{308}{L47}{1986}.

\bibitem{pac86}B. Paczy\'nski, \Journal{\it Ap. J. Lett.}{308}{L43}{1986}. 

\bibitem{mes94}P. M\'esz\'aros and M. Rees, 
\Journal{\MNRAS}{269}{L41}{1994}.

\bibitem{sar97}R. Sari and T. Piran, \Journal{\APJ}{485}{270}{1997}.

\bibitem{ree94}M. Rees and P. M\'esz\'aros, 
\Journal{\it Ap. J. Lett.}{430}{L93}{1994}. 

\bibitem{pap96}H. Papathanassiou and P. M\'esz\'aros, 
\Journal{\it Ap. J. Lett.}{471}{L91}{1996}.
 
\bibitem{mes97}P. M\'esz\'aros and M. Rees, 
\Journal{\it Ap. J.}{476}{232}{1997}.

\bibitem{fen93}E. E. Fenimore, \etal, 
\Journal{\it Nature}{366}{40}{1993}. 

\bibitem{hak96}J. Hakkila, \etal, \Journal{\APJ}{462}{125}{1996}.

\bibitem{sch92}B. E. Schaefer, in {\it Gamma-Ray Bursts:  Observations, 
Analyses and Theories}, eds. C.~Ho, R.~I.~Epstein and E.~E.~Fenimore 
(Cambridge Univ. Press, Cambridge, 1992), 107.

\bibitem{lar96}S. B. Larson, I. S. McLean, and E. E. Becklin, 
\Journal{\APJL}{460}{L95}{1996}.

\bibitem{lar97a}S. B. Larson and I. S. McLean,
\Journal{\APJ}{491}{in press}{1997}.

\bibitem{lar97b}S. B. Larson, \Journal{\APJ}{491}{in press}{1997}.

\bibitem{ban98}D.~L. Band and D. H. Hartmann, 
\Journal{\it Ap. J.}{\bf 493}{9709067}{1998}.

\bibitem{sch97}B. E. Schaefer, \etal, {\it Ap. J.}, astro-ph/9704278 (1997).

\bibitem{mad96}P. Madau, \etal, \Journal{\MNRAS}{283}{1388}{1996}.

\bibitem{wij97b}R.~A.~M.~J. Wijers, \etal,
{\it M.N.R.A.S.}, astro-ph/9708183 (1997).

\bibitem{tot97}T. Totani, \Journal{\it Ap. J. Lett.}{486}{L71}{1997}.

\bibitem{ban93}D. Band, \etal, \Journal{\APJ}{413}{281}{1993}. 

\bibitem{pir96}T. Piran and R. Narayan, in {\it
Gamma-Ray Bursts, AIP Conf. Proc. 384}, eds. C.~Kouveliotou, M.~F.
Briggs and G.~J.~Fishman (AIP, New York, 1996), 233. 

\bibitem{pre96}R. D. Preece, \etal,  
\Journal{\it Ap. J.}{473}{310}{1996}. 

\bibitem{str98}T. E. Strohmayer, \etal, {\it Ap. J.}, in press (1998).

\bibitem{din94}B. L. Dingus, \etal, in {\it Gamma-Ray Bursts, AIP Conf. Proc.
307}, eds. G.~J. Fishman, J.~J. Brainerd, and K.~Hurley (AIP, New
York, 1994), 22. 

\bibitem{hur94}K. Hurley, \etal, \Journal{\it Nature}{372}{652}{1994}.

\bibitem{gol83}S. V. Golenetskii, \etal, 
\Journal{\it Nature}{306}{451}{1983}. 

\bibitem{nor86}J. P. Norris, {\it et al.}, 
\Journal{\it Ap. J.}{301}{213}{1986}. 

\bibitem{kar94}V. E. Kargatis, \etal, 
\Journal{\APJ}{422}{260}{1994}.

\bibitem{for95}L. A. Ford, {\it et al.}, 
\Journal{\APJ}{439}{307}{1995}.

\bibitem{ban97b}D. L. Band, \Journal{\it Ap. J.}{486}{928}{1997}.

\bibitem{fen95}E. E. Fenimore, \etal, \Journal{\APJL}{448}{L101}{1995}.

\bibitem{lia96}E. Liang, and V. Kargatis, 
\Journal{\it Nature}{381}{49}{1996}.

\bibitem{maz82}E. P. Mazets, \etal, 
\Journal{\it Astrophys. Space Sci.}{84}{173}{1982}. 
 
\bibitem{hue87}G. J. Hueter, PhD Thesis (UC San Diego, 1987).

\bibitem{mur88}T. Murakami, \etal, \Journal{\it Nature}{335}{234}{1988}.
 
\bibitem{gra92}C. Graziani, \etal, in {\it Proc. Taos Workshop on 
Gamma-Ray Bursts}, eds. C.~Ho, R.~I.~Epstein, and
E.~E.~Fenimore (Cambridge Univ. Press, Cambridge, 1992), 407.

\bibitem{ale89}S. Alexander and P. M\'esz\'aros, 
\Journal{\it Ap. J. Lett.}{344}{L1}{1989}. 

\bibitem{wan89} J. C. L. Wang, \etal, \Journal{\PRL}{63}{1550}{1989}.

% (Paper~I)
\bibitem{pal94}D. Palmer, \etal, \Journal{\APJL}{433}{L77}{1994}.

% (Paper~IV)
\bibitem{ban96}D. Band, \etal, \Journal{\APJ}{458}{746}{1996}.

\bibitem{bri96b}M. S. Briggs, \etal, in {\it Gamma-Ray Bursts, AIP
Conf. Proc. 384}, eds. C.~Kouveliotou, M.~F. Briggs and G.~J.~Fishman
(AIP, New York, 1996), 153. 

% (Paper~II)
\bibitem{ban94}D. Band, \etal, \Journal{\APJ}{434}{560}{1994}.

% (Paper~III)
\bibitem{ban95}D. Band, \etal, \Journal{\APJ}{447}{289}{1995}.

% (Paper~V)
\bibitem{ban97}D. Band, \etal, \Journal{\APJ}{485}{747}{1997}.

\bibitem{fen93b}E. E. Fenimore, \etal, in {\it Compton
Gamma-Ray Observatory, AIP Conf. Proc. 280}, eds. M.~Friedlander,
N.~Gehrels, and D.~J.~Macomb (AIP, New York, 1993), 917. 

\end{thebibliography}
